\documentclass[
  journal=largetwo,
  manuscript=article-type,
  year=2020,
  volume=37,
]{cup-journal}

\usepackage{bm}

\usepackage{graphicx}

\usepackage{amsmath,amssymb}
\usepackage[nopatch]{microtype}
\usepackage{booktabs}
\usepackage{aas-macros}

\usepackage{hyperref}

\title{A long time ago in an LAE far, far away: a signpost of early reionisation or a nascent AGN at z = 13?}

\author{Joshua Cohon}
\affiliation{School of Earth and Space Exploration, Arizona State University, Tempe, AZ 85281, USA}
\email[Joshua Cohon]{jacohon@asu.edu}

\author{Christopher Cain}
\affiliation{School of Earth and Space Exploration, Arizona State University, Tempe, AZ 85281, USA}

\author{Rogier Windhorst}
\affiliation{School of Earth and Space Exploration, Arizona State University, Tempe, AZ 85281, USA}

\author{Anson D'Aloisio}
\affiliation{Department of Physics and Astronomy, University of California, Riverside, CA 92521, USA}

\author{Timothy Carleton}
\affiliation{School of Earth and Space Exploration, Arizona State University, Tempe, AZ 85281, USA}

\author{Yongda Zhu}
\affiliation{Steward Observatory, University of Arizona, 933 North Cherry Avenue, Tucson, AZ 85721, USA}


\begin{document}

\begin{abstract}

The JADES survey recently reported the discovery of JADES-GS-z13-1-LA at $z = 13$, the highest redshift Ly$\alpha$ emitter (LAE) ever observed. This observation suggests that either the intergalactic medium (IGM) surrounding JADES-GS-z13-1-LA is highly ionised, or the galaxy's intrinsic Ly$\alpha$ emission properties are extreme. We use radiative transfer simulations of reionisation that capture the distribution of ionised gas in the $z = 13$ IGM to investigate the implications of JADES-GS-z13-1-LA for reionisation. We find that if JADES-GS-z13-1-LA is a typical star forming galaxy (SFG) with properties characteristic of LAEs at $z \sim 6$, its detection suggests that the universe is $\gtrsim 5\%$ ionised by $z = 13$. We also investigate the possibility that the extreme properties of JADES-GS-z13-1-LA are driven by an AGN. Using a simple analysis based on the fact that AGN are expected to produce more ionising photons than SFGs, we estimate that the probability that JADES-GS-z13-1-LA hosts an AGN is $71\%$, $42\%$, and $15\%$ if the IGM is $< 1\%$, $\approx 5\%$ and $\approx 25\%$ ionised, respectively. We also highlight other features in the spectrum of JADES-GS-z13-1-LA that may be indicative of AGN activity, including strong Ly$\alpha$ damping wing absorption extending to $\sim 1300~\text{Å}$, and a possible CII*$\lambda1335$ emission line. Our findings strongly motivate dedicated follow-up observations of JADES-GS-z13-1-LA to determine whether it hosts an AGN.

\end{abstract}

\section{Introduction}
\label{sec:intro}

During cosmic reionisation, the neutral hydrogen in the intergalactic medium (IGM) was ionised by UV emission from the first galaxies. Despite substantial progress in the last decade, the timing of reionisation remains largely uncertain. Observations of the $z \lesssim 6$ Ly$\alpha$ forest of high-redshift quasars have localised the endpoint of reionisation to $z \approx 5.5$~\citep{Becker2015,Kulkarni2019,Keating2019,Nasir2020,Bosman2021,Zhu2022,Becker2024,Zhu2024,Spina2024,Qin2024b}. The Planck measurement of the CMB optical depth favours a reionisation midpoint in the redshift range of $\approx 7-8.5$~\citep{Planck2018,deBelsunce2021,Tristram2024}, and this picture is supported by constraints from quasar damping wings~\citep{Davies2018,Wang2020,Yang2020a} and Ly$\alpha$ emitting galaxies~\citep[LAEs,][]{Mason2018a,Mason2019}. However, the early stages, including when it started, remain largely un-constrained.  

Damping wing absorption of Ly$\alpha$ photons on the red side of line systemic in galaxy spectra have already been used to place some constraints on reionisation's early stages. This has been done using the visibility statistics of LAEs~\citep{Whitler2020,Wold2022,Morishita2023,Bruton2023,Tang2024b} and damping wing signatures in the spectra of Lyman-break galaxies~\citep[LBGs,][]{Bolan2022,Umeda2023,Kageura2025,Umeda2025,Mason2025,Huberty2025}. Indeed, these remain the only methods to date that have produced direct constraints on the reionisation history at $z > 9$\footnote{One exception is constraints placed on the ionisation fraction at $z \gtrsim 15$ by the CMB, although see~\citet{Wu2021} for a discussion of the limitations of this kind of analysis. In the future, 21 cm observations should provide better constraints on the ionisation fraction at this redshift \citep{Berkhout2024}}. Unfortunately, a dearth of statistical samples of spectra at these redshifts and modelling uncertainties has severely limited the precision of these constraints. 

Recently, the JADES survey \citep{Eisenstein2023} observed JADES-GS-z13-1-LA, which displays a modestly bright Ly$\alpha$ emission line with a rest-frame equivalent width (EW) of $42~\text{Å}$, at a remarkably high redshift - $z = 13$~\citep{Witstok2024}. This is well above the previous redshift record-holder for Ly$\alpha$ emission, GNz-11~\citep{Bunker2023}, and is at a redshift when the IGM is expected to be mostly neutral, and thus opaque to such emission~\citep{Mason2020,Hsiao2023,Nakane2023}. In a mostly neutral IGM at this redshift, the damping wing optical depth on the red side of Ly$\alpha$ should be sufficient to suppress emission by a factor of $10$ or more. Indeed,~\citet{Witstok2024} inferred the intrinsic EW of the JADES-GS-z13-1-LA emission line to be over $600~\text{Å}$ using an analytic model for the effect of the local IGM on its spectrum (see also~\citet{Qin2024}). 

JADES-GS-z13-1-LA is remarkable for two other reasons. The first is the presence of strong damping wing absorption {\it on the red side} of the Ly$\alpha$ emission line. This absorption is too strong to be from the IGM, and was interpreted\footnote{They also considered the possibility of a two-photon continuum, but found that the DLA scenario was a marginally better fit to the data. } by~\citet{Witstok2024} to be due to the presence of a nearby damped Ly$\alpha$ absorber (DLA). The combination of strong DLA absorption and Ly$\alpha$ emission is challenging to explain geometrically, although a few similar objects have been observed at lower redshifts~\citep[e.g.][]{Tacchella2025}. The second feature is the comparative faintness of the galaxy itself, which has $M_\text{UV}\approx-18.7$. This can be compared with two luminous spectroscopically confirmed high-redshift galaxies at $z\sim14$, JADES-GS-z14-0 and JADES-GS-z14-1, with $M_\text{UV}\approx-21,-19$ respectively~\citep{Carniani2024}, both of which lack Ly$\alpha$ emission (see also~\citet{Naidu2025} for another $z \approx 14$ example). 

These extreme properties raise questions about the origin of Ly$\alpha$ emission in JADES-GS-z13-1-LA. \citet{Witstok2024} suggests two possible scenarios: (i) a nuclear starburst driving Ly$\alpha$ emission by HII regions, which is scattered through a largely neutral ISM, or (ii) emission from an AGN viewed edge-on, resulting in damping wing absorption of the continuum by an accretion disk. This second possibility is of particular interest in light of ongoing debate about the importance of AGN in reionisation's earliest stages~\citep{Dsilva2023,Madau2024,Dayal2025}. Indeed, there are several lines of evidence that suggest that rapidly growing AGN may be ubiquitous at $z > 10$. These include the observed relationship between quasar lifetimes and supermassive black hole (SMBH) masses at high redshift~\citep{Yang2020a,Eilers2021,Jahnke2025}, and the observed over-abundance of bright galaxies at high redshifts~\citep{Hegde2024}. Several of these galaxies have already shown conclusive or tentative evidence of hosting AGN, including GNz-11~\citep{Maiolino2024}, UHZ1~\citep{Natarajan2024}, and GHZ2~\citep{Castellano2024}. These considerations motivate a more careful look at the possibility that JADES-GS-z13-1-LA hosts an AGN. 

Recently,~\citet{Qin2024} used numerical simulations of reionisation and a model for the statistics of $z = 13$ LAEs based on recent observations~\citep{Mason2018,Tang2024b} to assess the likelihood of observing JADES-GS-z13-1-LA. They found that, when accounting for redshift evolution in the intrinsic properties of bright LAEs within the first ionised bubbles, the chances of observing an LAE as bright as JADES-GS-z13-1-LA is at least a few percent. They found that the likelihood of observing JADES-GS-z13-1-LA is sensitive to both the global ionised fraction and the spatial morphology of the ionised regions hosting the LAEs. Their work further motivates investigations of whether (and under what conditions) JADES-GS-z13-1-LA requires an early start to reionisation, a question we take up here. 

In this work, we present new modelling of the emission spectrum of JADES-GS-z13-1-LA informed by radiative transfer (RT) simulations of reionisation. We focus on two closely related questions: (1) what are the implications of JADES-GS-z13-1-LA for the timing of reionisation? and (2) is the Ly$\alpha$ emission from JADES-GS-z13-1-LA powered by an AGN? This work is outlined as follows. In \S\ref{sec:Lya}, we discuss the methods we use to model reionisation and the spectrum of JADES-GS-z13-1-LA. In \S\ref{sec:inference}, we discuss the intrinsic emission properties we infer from our modelling. In \S\ref{sec:likelihood}, we discuss the probability that this galaxy hosts an AGN, and how that couples to the reionisation history. Throughout this paper, we assume the following cosmological parameters: $\Omega_m = 0.305$, $\Omega_\Lambda=1-\Omega_m$, $\Omega_b = 0.048$, $h= 0.68$, $n_s = 0.96$ and $\sigma_8 = 0.82$, consistent with the results from~\citet{Planck2018}.

\section{Modelling JADES-GS-z13-1-LA}
\label{sec:Lya}

This work uses numerical simulations to model the state of the universe at $z=13$, and Bayesian methods to model the properties of JADES-GS-z13-1-LA. In this section, we describe the technical details of our models.

\subsection{Simulations of reionisation}
\label{subsec:FlexRT}

\begin{figure*}[hbt!]
    \centering
    \includegraphics[scale=0.165]{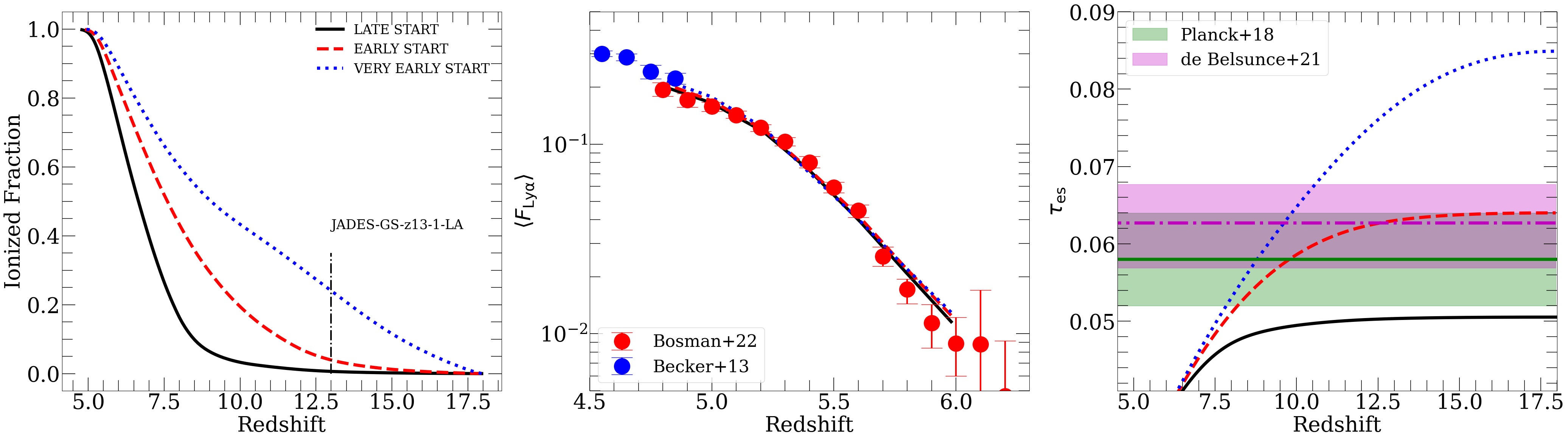}
    \caption{Properties of the reionisation models used in this work. {\bf Left:} the volume-weighted mean ionised fraction vs. redshift (reionisation history), with the redshift of JADES-GS-z13-1-LA indicated. The \textsc{late start}, \textsc{early start}, and \textsc{very early start} models have ionised fractions of $< 1\%$, $\approx 5\%$, and $\approx 25\%$ at $z = 13$, respectively. 
    {\bf Middle:} mean transmission of the Ly$\alpha$ forest at $4.8 < z < 6$ compared to measurements from~\citet{Becker2013} and~\citet{Bosman2021}. {\bf Right:} CMB electron scattering optical depth. The \textsc{late start} model is slightly more than $1\sigma$ below the Planck measurement, while the \textsc{early start} case is within $1\sigma$ of the fiducial measurement and the re-analysis by~\citet{deBelsunce2021}. The \textsc{very early start} case is in more than $4\sigma$ tension with the fiducial Planck result.  }
    \label{fig:compare_models}
\end{figure*}

We model transmission through the IGM using radiative transfer simulations of reionisation run with FlexRT, the radiative transfer (RT) code described in~\citet{Cain2024c}. Our simulation setup and approach is the same as that described in~\citet{Cain2024b} - we refer the reader to \S3 of that work for details, and summarise salient aspects here. In FlexRT, the redshift evolution of the ionising photon emissivity from all galaxies, $\dot{N}_{\gamma}$, is free to be adjusted at all redshifts to obtain a desired reionisation history and/or calibrate the simulation to match one or more observables~\citep[see also e.g.][]{Kulkarni2019,Asthana2024}. We divide $\dot{N}_{\gamma}$ between halos in the simulation by assuming that the ionising output of an individual halo is proportional to its UV luminosity, $\dot{N}_{\gamma} \propto L_{\rm UV}$. We assign UV luminosities to halos by abundance-matching to the UV luminosity function measured by~\citet{Adams2023}. In this work, we calibrate $\dot{N}_{\gamma}$ at $z \lesssim 7$ to match the mean transmission of the Ly$\alpha$ forest at $z \leq 6$ measured by~\citet{Bosman2021} (see \S3.2 of~\cite{Cain2024b}). 

We consider three reionisation histories, all of which complete reionisation at $z \approx 5-5.5$, as required by our calibration to the Ly$\alpha$ forest. The models differ significantly in their early stages - that is, when reionisation starts and how quickly it progresses. We use the \textsc{late start/late end} and \textsc{early start/late end} models studied in~\cite{Cain2024b}, alongside a third model which starts reionisation even earlier than the latter. Since all our models end reionisation late, for brevity we refer to these as the \textsc{late start}, \textsc{early start}, and \textsc{very early start} models, respectively\footnote{Note that these three models are similar to the three reionisation histories studied in~\citet{Asthana2024}. }. In Figure~\ref{fig:compare_models}, we show the reionisation history (left), the mean Ly$\alpha$ forest transmission, $F_{\text{Ly}\alpha}$, at $4.8 \leq z \leq 6$ (middle), and the CMB electron scattering optical depth (right). We denote the redshift of JADES-GS-z13-1-LA in the left panel. The ionised fraction in the \textsc{late start}, \textsc{early start}, and \textsc{very early start} at $z = 13$ is $< 1\%$, $\approx 5\%$, and $\approx 25\%$. These scenarios differ considerably in their Ly$\alpha$ transmission properties around star-forming halos at $z \gtrsim 10$, despite all being calibrated to match the same Ly$\alpha$ forest transmission properties at low redshift.

\subsection{Modelling Ly$\alpha$ transmission statistics}
\label{subsec:LyaT}

We follow the procedure described in \S3.3 of~\citet{Cain2024b} to model Ly$\alpha$ transmission through the IGM on the red side of line systemic. We run an Eulerian hydrodynamics simulation of the IGM using the RadHydro code of~\citet{Trac2004,Trac2006} with the same initial conditions used in the dark matter N-body simulation used to generate the halos used in the RT simulations. This run has a box size of $200$ $h^{-1}$Mpc and $N = 2048^3$ uni-grid gas cells, for a spatial resolution of $\Delta x = 97$ $h^{-1}$kpc. While this resolution is insufficient to capture the physics at play in setting the properties of the emerging Ly$\alpha$ line and absorption by the CGM of galaxies, it is enough to capture the effect of large-scale gravitational inflows around halos and local density fluctuations in the IGM around galaxies. The former play a particularly important role in setting Ly$\alpha$ transmission near the line centre~\citep{Park2021}. 

We model the Ly$\alpha$ absorption line profile using the analytic approximation given in~\citet{TepperGarcia2006}. At $z = 13$, we trace $50$ randomly oriented sightlines around halos with $-19<M_\text{UV}<-18$, for a total of $\sim50,000$ sightlines. We integrate a distance of $200$ $h^{-1}$Mpc away from the halos, sufficient to converge on the damping wing absorption on the red side of systemic in the neutral IGM. To avoid including absorption arising from within the un-resolved halos themselves, we set the start of each sightline $500$ $h^{-1}$kpc away from the halo centre. Because the spatial resolution of the simulation is too low to capture the integration over the central line profile in regions with high inflow velocities, we artificially boost the spatial resolution in the line integration by a factor of $4\times$, and use a cloud-in-cell scheme to interpolate grid quantities at intermediate points\footnote{This is unimportant for modelling damping wing absorption itself, but is necessary to avoid spurious transmission spikes in cases where line-centre absorption is red-shifted by a large-scale inflow. 
 }~\citep[as described in][]{Cain2024b,Gangolli2024}. We find this procedure produces converged transmission spectra in nearly all cases. 

\begin{figure}[hbt!]
    \centering
    \includegraphics[scale=0.6]{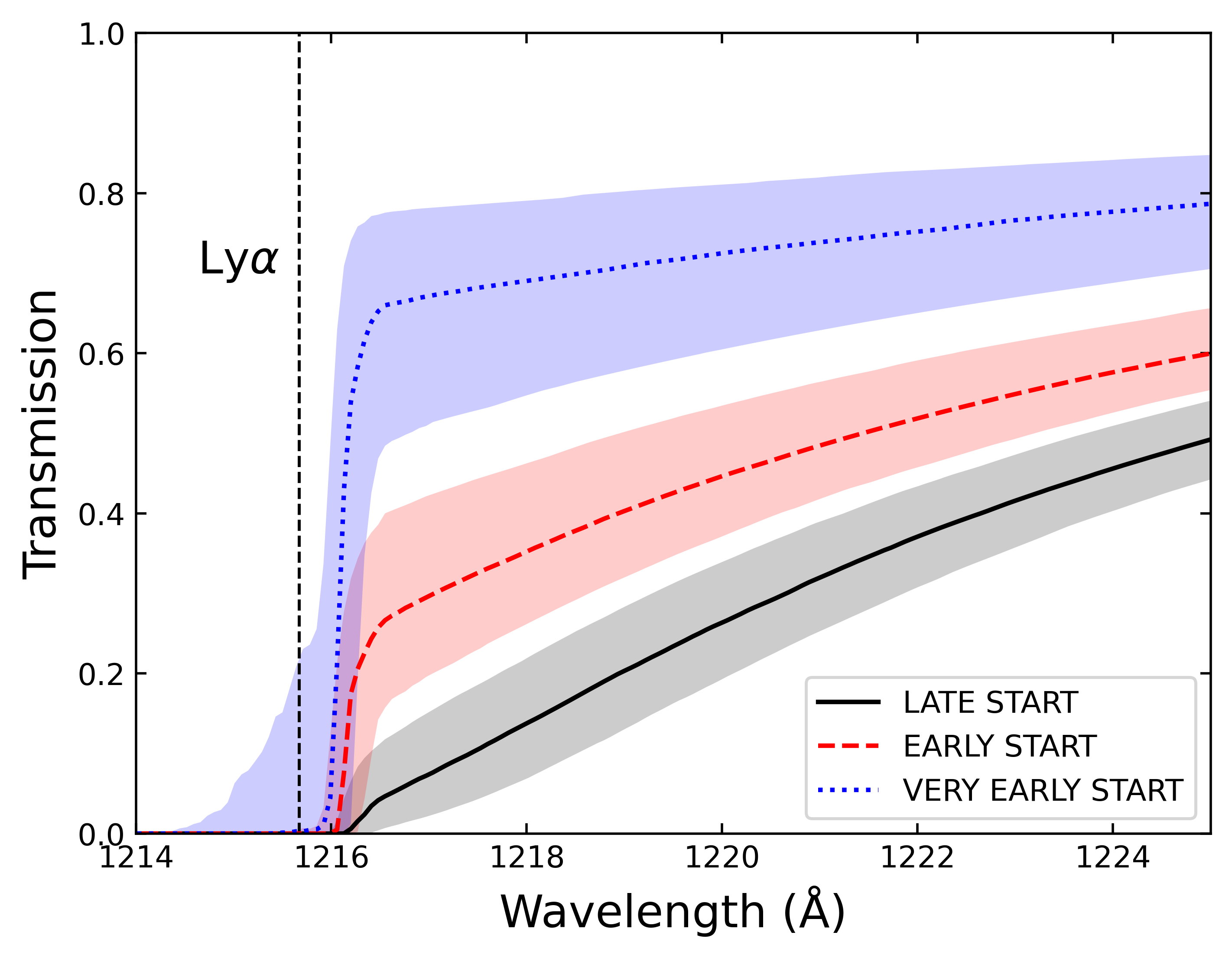}
    \caption{Median Ly$\alpha$ transmission as a function of wavelength in each of our reionisation models. The colours and line styles are the same as in Figure~\ref{fig:compare_models}, and the shaded regions denote $1\sigma$ sightline-to-sightline scatter. We see a substantial difference in IGM damping wing transmission between reionisation models, especially close to systemic Ly$\alpha$.  }
    \label{fig:transmission}
\end{figure}

In Figure~\ref{fig:transmission}, we show the median IGM transmission ($T_{\rm IGM}$) as a function of wavelength ($\lambda$) for each of our reionisation models, using the same line styles as Figure~\ref{fig:compare_models}. The shaded regions indicate $1\sigma$ sightline-to-sightline scatter. The vertical dashed line denotes the systemic Ly$\alpha$ wavelength. We see a marked difference between models, especially close to systemic. The \textsc{late start} model has $T_{\text{IGM}} < 20\%$ within a few $\text{Å}$ of systemic, and declines gradually to $0$ at line centre. In the \textsc{early start} and \textsc{very early start} models, $T_{\rm IGM}$ jumps to $30\%$ and $60\%$ close to systemic, respectively, reflecting the presence of ionised bubbles around most halos in those models. We thus expect required intrinsic Ly$\alpha$ brightness to produce a given observed brightness to differ considerably between these scenarios, especially for lines emitted close to systemic. 

\subsection{Spectrum model}
\label{subsec:spectrum}

The available NIRSpec/PRISM spectrum for JADES-GS-z13-1-LA shows only one clear emission line — Ly$\alpha$ — with no other lines conclusively detected. SED fitting of JADES-GS-z13-1-LA indicates a young, metal-poor stellar population; however, standard SED fitting codes cannot model the attenuated Ly$\alpha$ emission line and strong Ly$\alpha$ damping wing that we observe in this object's spectrum, so we use the phenomenological model described in~\citet{Witstok2024}.

The spectrum of JADES-GS-z13-1-LA also has an unusual UV continuum turnover around $\lambda=1335~\text{Å}$. We model this using a DLA, as discussed above. However, if there were a DLA in front of the Ly$\alpha$ emission source, we expect that the DLA would completely attenuate the Ly$\alpha$ emission line, so the Ly$\alpha$ emission source cannot be in the same place as the UV continuum source. \citet{Witstok2024} suggests two morphological models of JADES-GS-z13-1-LA in which the continuum is attenuated by a DLA, but the emission line is not: Ly$\alpha$ photons escape either through ionisation cones or diffusion through an inhomogeneous ISM. It is also possible that the Ly$\alpha$ photons diffused directly through the DLA and were redshifted in the process, as described in~\citet{Dijkstra2014}, but our calculations indicate that this would redshift the Ly$\alpha$ line significantly more than we observe.

Following~\citet{Witstok2024}, we model the spectrum as a Gaussian intrinsic Ly$\alpha$ emission line plus a power law UV continuum,\footnote{\citet{Witstok2024} finds that a power-law continuum fits the spectrum better than a $2\gamma$ continuum.} with the continuum subject to DLA absorption and the whole spectrum subject to IGM absorption. The DLA transmission, $T_{\text{DLA}}(\lambda)$, is modelled using the analytic fit prescribed in~\citet{TepperGarcia2006} with the HI column density $N_{\rm HI}$ as a free parameter. The IGM transmission, $T_{\text{IGM}}(\lambda)$, is extracted from the simulation as described in \S\ref{subsec:LyaT}. We parameterise the continuum flux using two free parameters: a characteristic wavelength $\lambda^*$ and the UV slope $\beta_{\text{UV}}$:
\begin{equation}
\label{eq:Fcont}
F_{\text{cont}}(\lambda) = \Big(\frac{\lambda}{\lambda^*}\Big)^{\beta_{\text{UV}}}~10^{-21}~\text{erg}~\text{s}^{-1} \text{cm}^{-2}~\text{Å}^{-1}.
\end{equation}

We parameterise the intrinsic emission line using three free parameters: the velocity offset $\Delta{v}$, the intrinsic equivalent width $\text{EW}$, and the velocity width $\sigma$. We set the line amplitude so that the line has equivalent width $\text{EW}$. Using these variables, the line flux $F_{\text{Ly}\alpha}(\lambda)$ is given by a Gaussian profile centred at $\lambda_{\text{Ly}\alpha}+\Delta\lambda$ with standard deviation $\sigma_\lambda$, where $\Delta \lambda$, $\sigma_\lambda$ are the wavelength-space equivalents of $\Delta v$, $\sigma$ respectively. The observed spectrum is modelled by:
\begin{equation} 
\label{eq:Fobs}
F_{\text{obs}}((z+1)\lambda) = \Big[F_{\text{Ly$\alpha$}}(\lambda) + F_{\text{cont}}(\lambda)T_{\text{DLA}}(\lambda)\Big]T_{\text{IGM}}(\lambda).
\end{equation}
where the redshift $z$ is taken as a free parameter since we have no other emission lines with which to precisely constrain it. The observed spectrum is convolved with the JWST NIRSpec PRISM broadening kernel as reported in~\citet{marshall2025} to obtain the final modelled observations. We use the dynamic nested sampler from the \textsc{dynesty} package to compute posterior distributions of parameters using the random walk method\footnote{The posteriors we get from dynamic nested sampling are compatible with posteriors we get from a Monte Carlo Markov-chain (MCMC) analysis using the Metropolis-Hastings sampler from the \textsc{emcee} package, so we conclude that our results are robust to the sampling method used.}~\citep{Skilling2004, Speagle2020}. We assume a standard Gaussian likelihood to handle measured uncertainties in the observed spectrum. 

\begin{figure}[h!]
    \includegraphics[scale=.59]{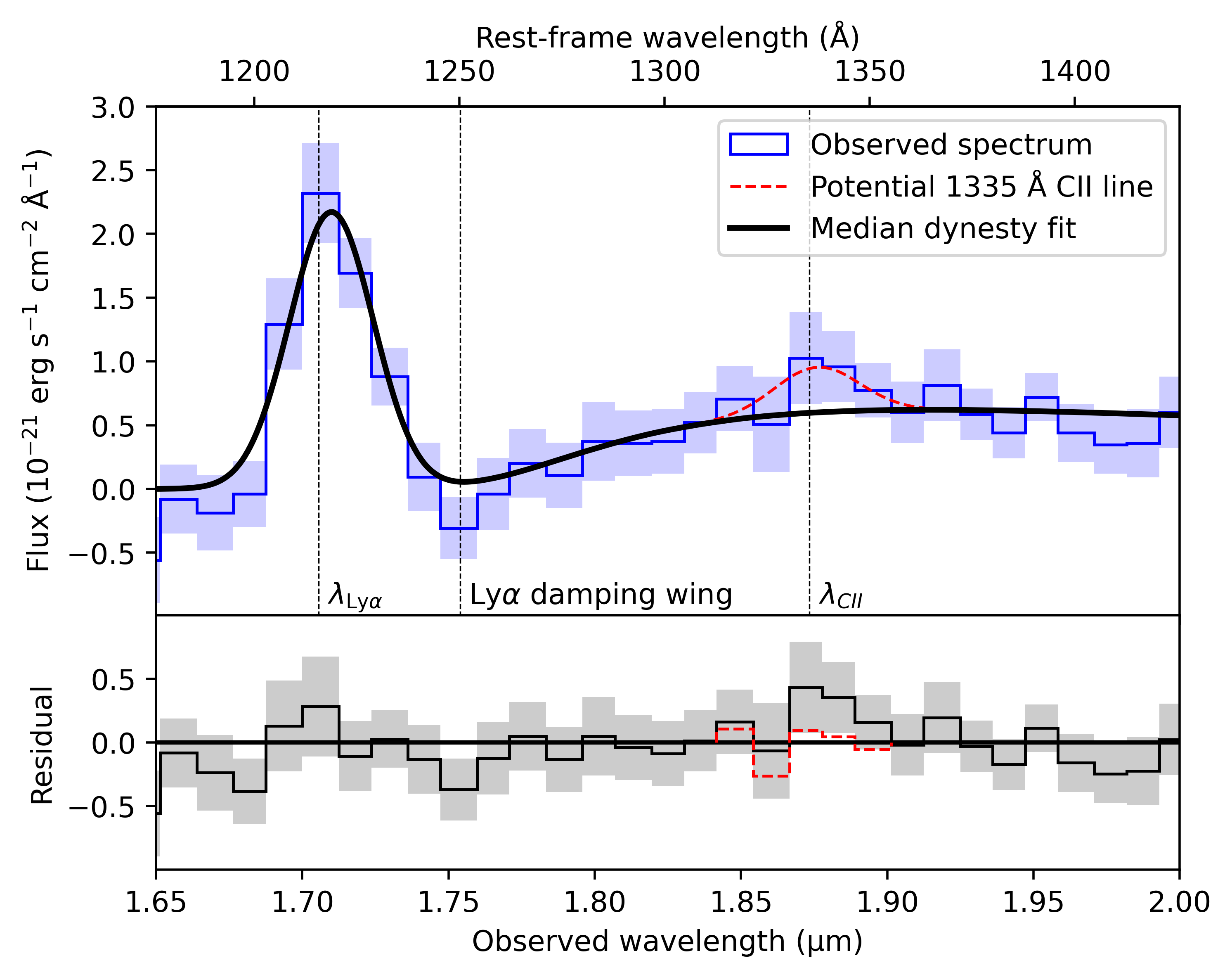}
\caption{\textbf{Top:} An example fit of the spectrum model for the \textsc{early start} model, obtained from the median of the highest-likelihood fits from each sightline's \textsc{dynesty} posterior distribution for the \textsc{early start} reionisation model. The blue histogram indicates the PRISM SED from~\citet{Witstok2024}, to which we fit the model in our analysis. The shaded regions indicate $1\sigma$ uncertainties for each spectral bin. The black line indicates the model, and the red dashed line indicates the model with an additional CII* emission line with $\text{EW}_\text{CII*}=15~\text{Å}$. \textbf{Bottom:} residuals from the fit. The shaded regions indicate the same $1\sigma$ uncertainties as above. The red dashed histogram indicates residuals when the potential CII* line is included in the model. Note that the residuals are consistent with continuum fluctuations except around $\lambda=1335~\text{Å}$. Including the CII* emission line smooths out the residuals and makes them consistent with continuum fluctuations.}
    \label{fig:fit}
\end{figure}

In Figure~\ref{fig:fit}, we show the spectrum of JADES-GS-z13-1-LA (blue line), including $1\sigma$ uncertainties denoted by the shaded regions. The peak on the left is the Ly$\alpha$ emission line, and the recovery of the flux to the right is well-fit by a damping wing profile (see annotations). The black curve shows the maximum-likelihood fit assuming the \textsc{early start} reionisation model. The bottom panel shows the residuals of the fit. We see that, as in~\cite{Witstok2024}, Eqns.~\ref{eq:Fcont}-\ref{eq:Fobs} capture well the essential features of the spectrum near Ly$\alpha$. Note that the UV continuum parameters ($\lambda^*$, $\beta_\text{UV}$, $N_\text{HI}$) are constrained by spectral features significantly redward of Ly$\alpha$ systemic. The power-law continuum parameters $\lambda^*$, $\beta_\text{UV}$ are constrained by flux at wavelengths with $\lambda\gtrsim 1335~\text{Å}$ redward of the damping wing absorption, and the DLA column density is constrained by flux at wavelengths with $1250~\text{Å}\lesssim \lambda \lesssim 1335~\text{Å}$, where the DLA absorption dominates over that from the IGM. Therefore, they do not vary significantly when we change the IGM transmission curve. 

One notable deviation between the spectrum of JADES-GS-z13-1-LA and our fiducial best-fit is the elevated flux at the edge of the damping wing feature, around $\lambda = 1335$ Å. Because its width is exactly what we would expect from a sharp emission feature of negligible spread broadened by the NIRSpec PRISM, we posit that some of the flux may be from a CII*$\lambda1335$ emission line.\footnote{The CII*$\lambda1335$ doublet line has been detected as early as $z=11$ in the spectrum of GN-z11~\citep{Maiolino2024}, so it could plausibly appear at $z=13$.} The red-dashed curve is a second fit to the spectrum that includes this line, and we see that it fits this spectral feature well. The inferred rest-frame EW for the CII* line is $15~\text{Å}$, with a velocity offset $\Delta v_\text{CII*}$ that varies somewhat with reionisation history\footnote{This occurs because we infer slightly different redshifts for each reionisation history. The average offset we infer is $\Delta v_\text{CII*}\lesssim150~\text{km/s}$.}. We find that although including a CII*$\lambda1335$ emission line produces somewhat higher likelihoods, it only negligibly influences the other parameters, so the CII* line can be safely ignored when modelling the Ly$\alpha$ emission line and DLA damping wing feature. An important caveat is that that the PRISM spectrum of JADES-GS-z13-1-LA does not similarly suggest possible [CII]$\lambda2326$ or CIII]$\lambda1909$ emission lines, which we would expect to see if we observe a CII*$\lambda1335$ emission line in an AGN spectrum~\citep{Moy2002,Humphrey2014}. Further, we estimate the peak signal-to-noise of the inferred line (if it is there) to be close to unity, so we do not claim a statistically significant detection.  

\section{Inference on the properties of JADES-GS-z13-1-LA}
\label{sec:inference}

\begin{table*}[hbtp!]

    \centering
    \begin{tabular}{c | c c c c c c c}
        & $z$ & $\lambda^*$ (Å) & $\beta_\text{UV}$ & $N_\text{HI}$ ($10^{22}~\text{cm}^{-2}$) & $\Delta v$ (km/s) & $\text{EW}$ (Å) & $\sigma$ (km/s) \\
        \hline
        Prior interval & $[12.96,13.10]$ & $[1000,1400]$ & $[-6,-2]$ & $[1.0,20.0]$ & $[0,500]$ & $[0,2000]$ & $[200,650]$ \\
        \hline
        \textsc{late start} & $13.02^{+0.02}_{-0.02}$ & $1304^{+37}_{-49}$ & $-4.4^{+0.7}_{-0.7}$ & $12.8^{+4.0}_{-3.8}$ & $349^{+112}_{-181}$ & $1443^{+401}_{-549}$ & $520^{+97}_{-164}$ \\
        \textsc{early start} & $13.03^{+0.02}_{-0.02}$ & $1288^{+40}_{-53}$ & $-4.2^{+0.7}_{-0.7}$ & $11.7^{+4.1}_{-3.5}$ & $217^{+180}_{-151}$ & $726^{+701}_{-410}$ & $415^{+154}_{-145}$ \\ 
        \textsc{very early start} & $13.04^{+0.02}_{-0.02}$ & $1281^{+42}_{-55}$ & $-4.1^{+0.7}_{-0.7}$ & $11.4^{+4.1}_{-3.5}$ & $194^{+193}_{-141}$ & $161^{+166}_{-65}$ & $414^{+157}_{-147}$ 
        
    \end{tabular}
    
    \caption{Priors and posteriors for all parameters from \textsc{dynesty} modelling of JADES-GS-z13-1-LA. We give results for all three of our reionisation histories.  Each prior is a uniform prior over the given interval. Note that only $\Delta v$ and $\text{EW}$ change significantly with reionisation history.}
    \label{table:params}
    
\end{table*}

\begin{figure*}[hbtp!]
    \begin{tabular}{cc}
    (a) \textsc{late start} & (b) \textsc{late start} with wider $\Delta v$ prior \\
    \includegraphics[scale=.6]{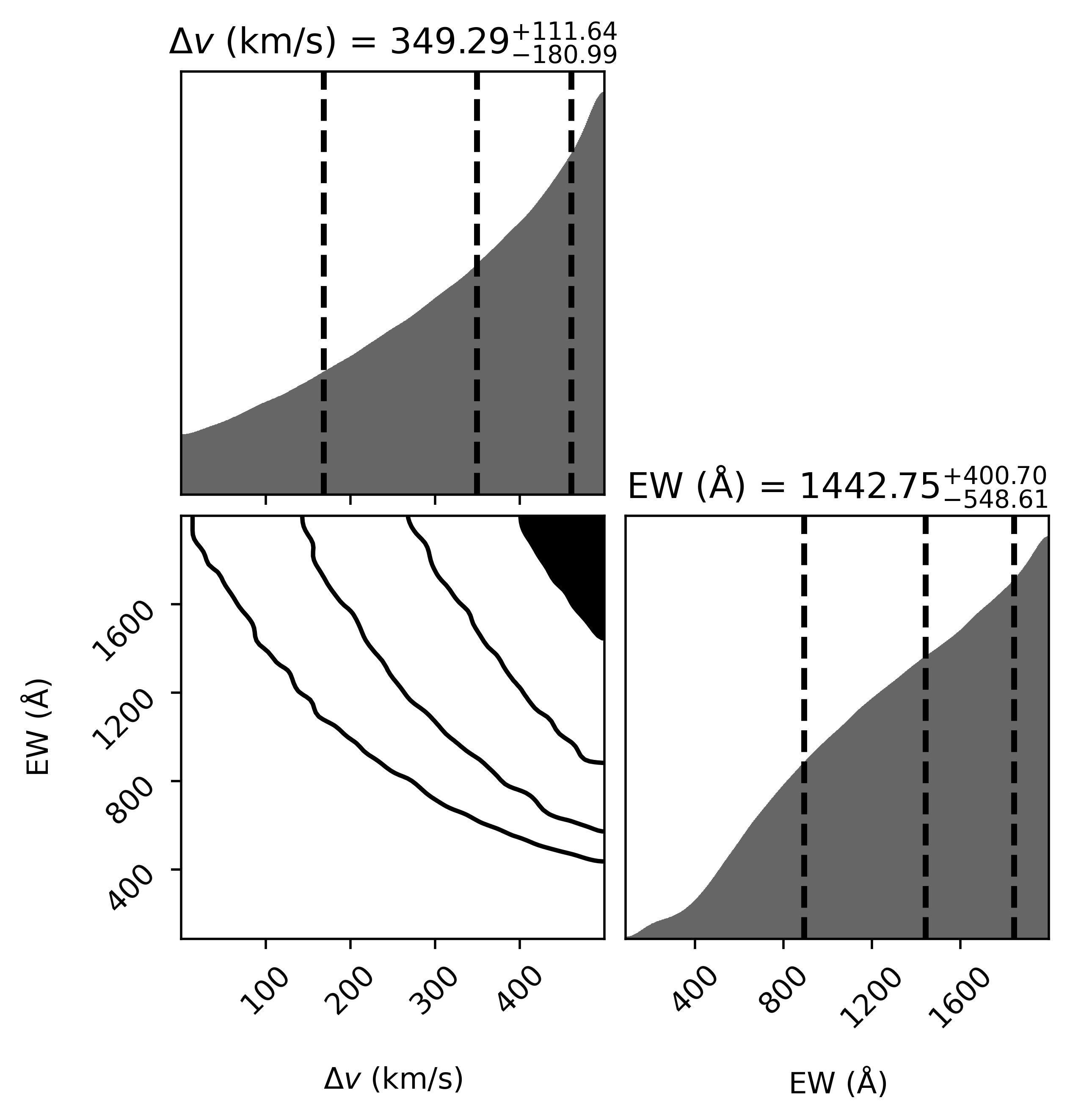} &  \includegraphics[scale=.6]{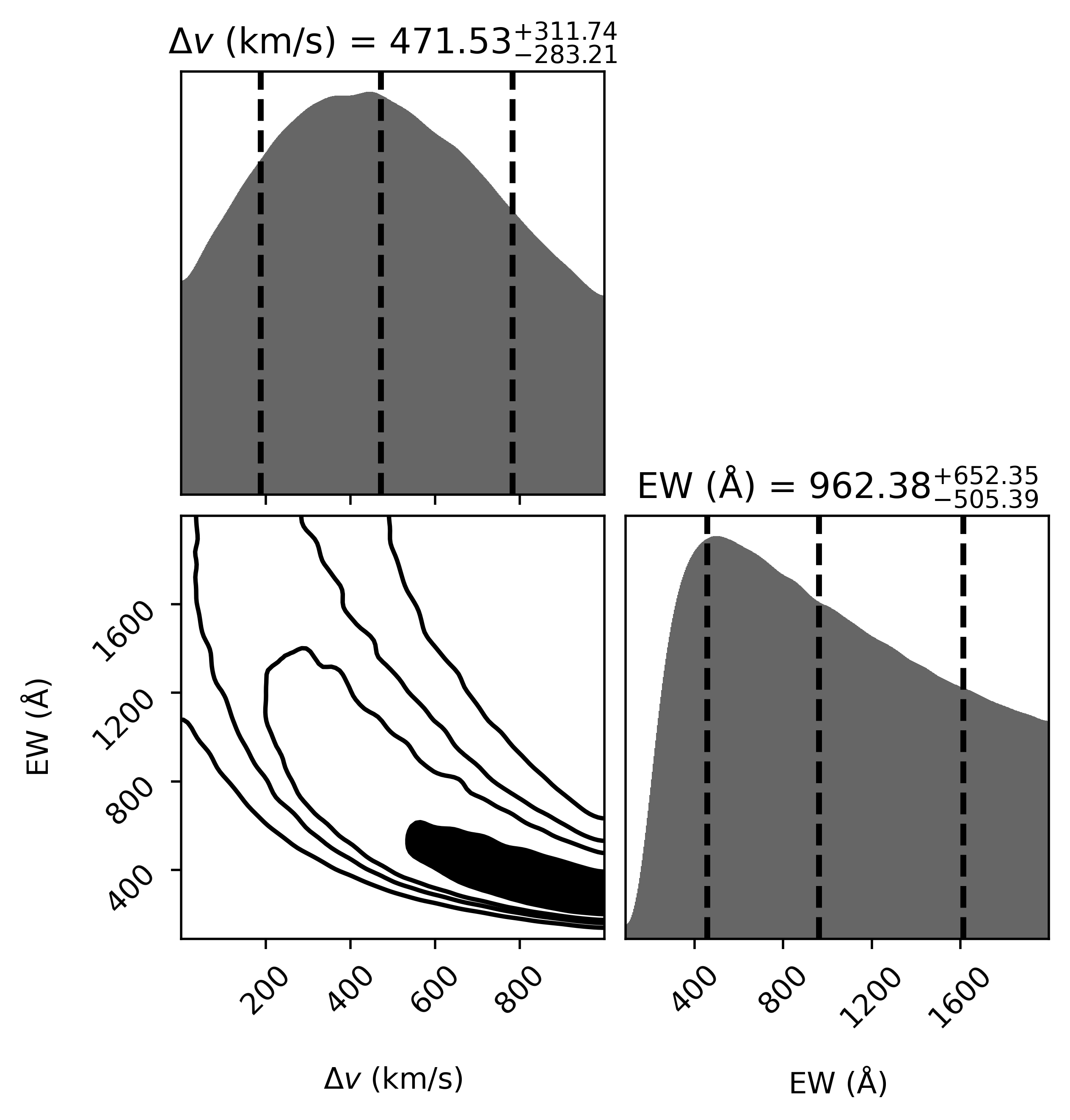} \\
    \includegraphics[scale=.6]{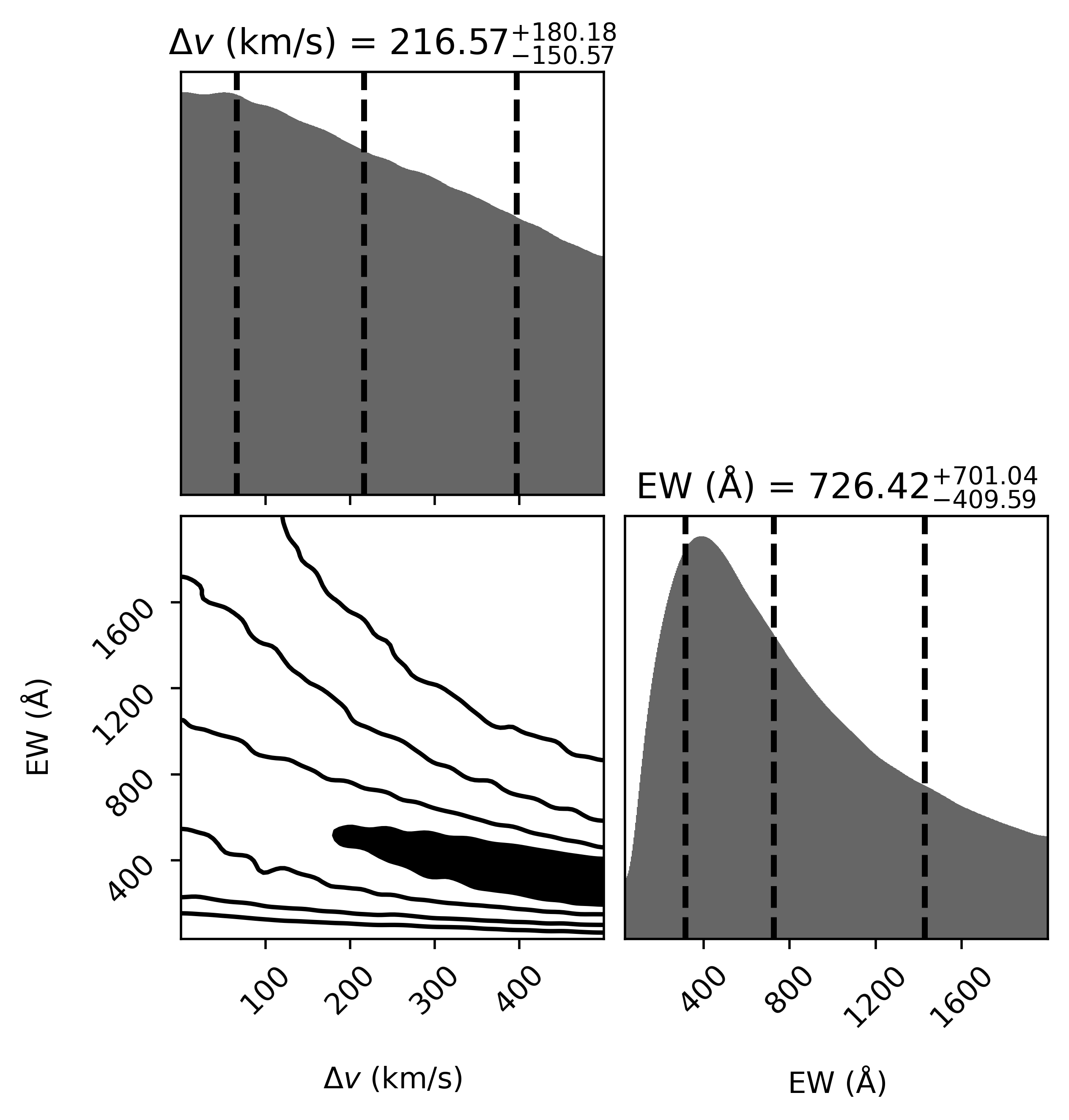} &  \includegraphics[scale=.6]{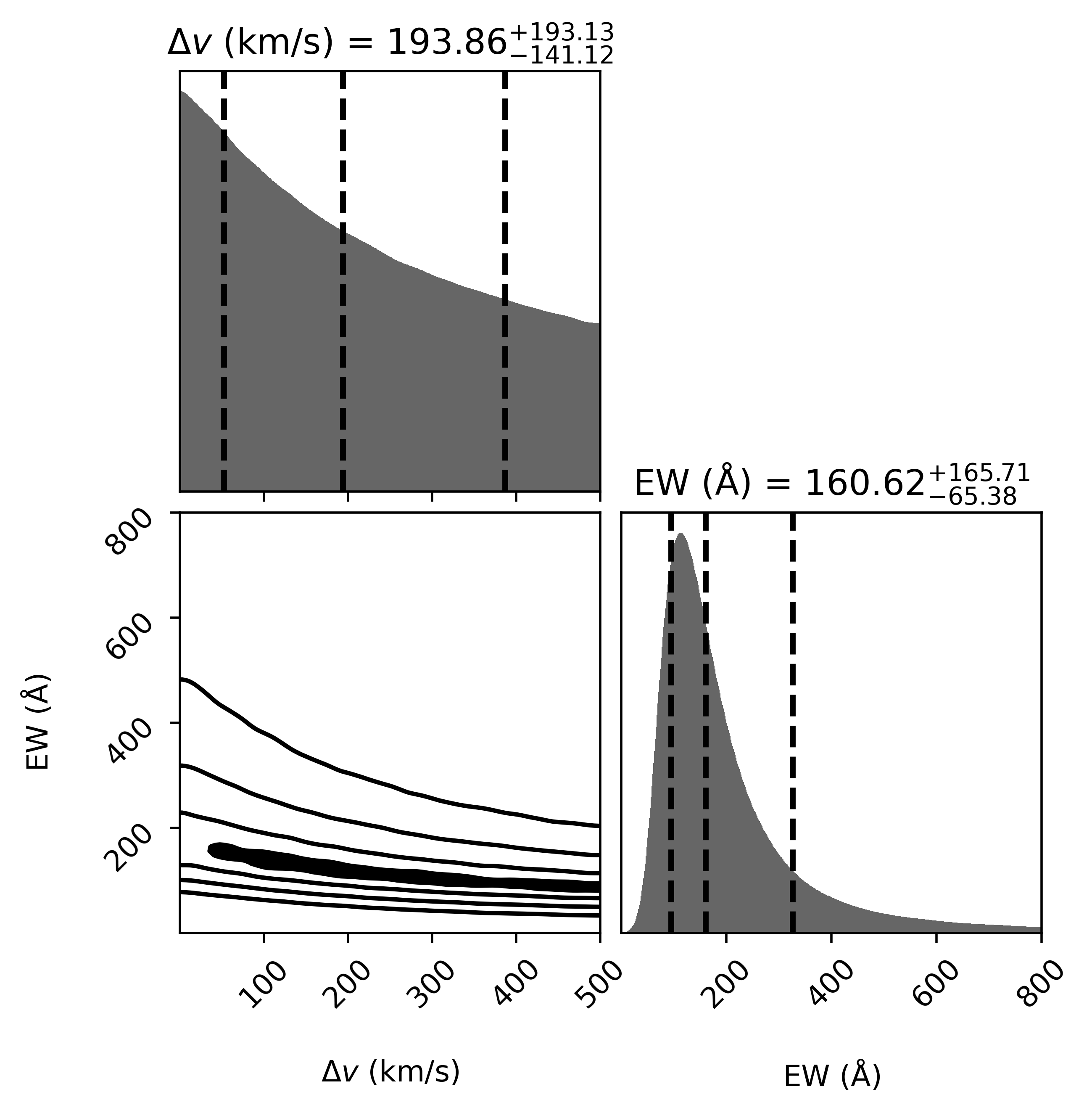} \\
    (c) \textsc{early start} & (d) \textsc{very early start}
    \end{tabular}
\caption{Velocity offset and intrinsic equivalent width joint posteriors for all three reionisation histories. In each plot, the upper panel shows the $\Delta v$ PDF, the lower-left panel shows the $0.5\sigma$, $1\sigma$, $1.5\sigma$, and $2\sigma$ contours for the $\Delta v$-$\text{EW}$ joint PDF, and the lower-right panel shows the $\text{EW}$ PDF. \textbf{(a)} shows the posterior for \textsc{late start}, which favours high $\Delta v$ and $\text{EW}$. \textbf{(b)} shows the posterior for \textsc{late start} when we use a uniform $\Delta v$ prior on the interval $[0,1000]$ rather than $[0,500]$. We include this to show that the \textsc{late start} model favours $\Delta v \approx 500~\text{km/s}$ if we allow $\Delta v$ to go out to $1000~\text{km/s}$. In subsequent calculations, we use the posterior shown in (a). Note that the $\Delta v$ axis is rescaled in this plot. \textbf{(c)} shows the posterior for \textsc{early start}, which peaks around $\text{EW}\approx400~\text{Å}$ and favours lower $\Delta v$. \textbf{(d)} shows the posterior for \textsc{very early start}, which peaks around $\text{EW}\approx100~\text{Å}$ and favours lower $\Delta v$. Note that the $\text{EW}$ axis is rescaled in this plot.
    }
    \label{fig:smallposteriors}
\end{figure*}

To determine the implications of JADES-GS-z13-1-LA for the history of reionisation, we performed a dynamic nested sampling of the spectrum parameters with each of the reionisation models described in \S\ref{sec:Lya}. To account for sightline-to-sightline variation in IGM transmission, we performed the modelling for 100 randomly chosen sightlines from dark matter halos with $-19\leq M_{\text{UV}}\leq -18$, and obtained the overall posterior distribution by averaging the posteriors from each sightline. We have checked that 100 sightlines is sufficient for a well-converged posterior. Note that this averaging procedure is justified by the fact that all sightlines are equally likely to be observed.

In Table \ref{table:params}, we give the priors assumed on each of our parameters (top row) and the median and $1\sigma$ ranges for each parameter in of our reionisation models. Our priors are uniform and make few assumptions about the intrinsic properties of JADES-GS-z13-1-LA. We get well-converged distributions for $z$, $\lambda^*$, $\beta_{\text{UV}}$, and $N_{\text{HI}}$ with any reasonable prior. The $\Delta v$ prior follows \citet{Witstok2024} and delineates the range of velocity offsets that one typically sees in a galaxy of this magnitude (see Figure 2 of~\citet{Qin2024} and associated references). The $\text{EW}$ prior is wide because of the extreme nature of this galaxy, allowing for an $\text{EW}$ several times higher than typically seen in lower-redshift LAEs. The $\sigma$ prior follows that used in \citet{Witstok2024} - note that we are unable to meaningfully constrain this parameter due to instrumental broadening the PRISM spectrum. We see that $z$, $\lambda^\ast$, $\beta_{\rm UV}$, and $N_{\rm HI}$ are reasonably well-constrained relative to the prior range and are reasonably insensitive to the reionisation history. The velocity offset varies mildly with reionisation history, while the EW varies strongly - by nearly an order of magnitude between the \textsc{late start} and \textsc{very early start} models. Throughout the rest of this work, our main focus is on the $\text{EW}$ and $\Delta v$ parameters of the Ly$\alpha$ emission line, as these most sensitive to the IGM neutral fraction. 

In Figure~\ref{fig:smallposteriors}, we show the posteriors on EW and $\Delta v$ for each of our reionisation histories. In each panel, we show the $1$D posteriors of $\Delta v$ and EW on the diagonal, and the joint posterior in the lower left. The dark shaded region and lines in the joint posterior show the $0.5\sigma$, $1\sigma$, $1.5\sigma$, and $2\sigma$ ranges. Panel (a) shows results for the \textsc{late start} reionisation history assuming our fiducial priors on $\Delta v$ and EW given in Table \ref{table:params}, Panel (b) shows the same results, but assuming a somewhat wider $\Delta v$ prior of $[0,1000]$ km/s. Panels (c) and (d) show our fiducial results for the \textsc{early start} and \textsc{very early start} models, respectively. 

We see in panel (a) that the \textsc{late start} model prefers high values for the Ly$\alpha$ emission parameters, with $\text{EW}\approx 1400 \text{Å}$ and $\Delta v \approx 350$ km/s, respectively. This model requires a very bright and highly redshifted intrinsic line because of the large damping wing optical depth near line centre in a $< 1\%$ ionised IGM (see Figure~\ref{fig:transmission}). In panel (b), we allow a wider prior on $\Delta v$ and see that a value of $\approx 500$ km/s is preferred by the data. For galaxies of UV magnitude similar to that of JADES-GS-z13-1-LA, this velocity offset is on the high end of what is observed at lower redshifts~\citep{Erb2014}. It is also well above the expected maximum velocity dispersion of a galaxy with this UV magnitude, which is around $140~\text{km/s}$.\footnote{This was derived using equations in \S4.2 of \citet{Navarro1995} and our simulation's $z=13$ mass-luminosity relations.} Note that resonant scattering could also broaden the emission line (see \cite{Dijkstra2014}).

We also see a clear degeneracy between $\Delta v$ and EW in the joint posterior - larger $\Delta v$ requires smaller intrinsic EW, and vice versa. This is because an intrinsic line redshifted further from systemic would encounter a smaller damping wing optical depth, thus requiring a lower intrinsic brightness. The large scatter in the 2D posterior comes comparably from both (i) large scatter in $T_{\rm IGM}$ between different sightlines, reflecting our lack of constraints on IGM attenuation in the immediate vicinity of JADES-GS-z13-1-LA, and (ii) observational uncertainty regarding the spectrum of JADES-GS-z13-1-LA. Even with a wide prior on $\Delta v$, the best-fit EW is nearly $1000~\text{Å}$, on the extreme high end of observations for lower-redshift samples~\citep{Erb2014,Steidel2014}. With a restricted $\Delta v$ prior, the EW posterior peaks at the maximum allowed value, $2000~\text{Å}$.

We find progressively less extreme inferred properties in the \textsc{early start} and \textsc{very early start} cases. In both cases, the preferred $\Delta v$ is $\approx 200$ km/s, and the posteriors are very broad. The \textsc{very early start} case even seems to mildly prefer $\Delta v = 0$. In the \textsc{early start} model, the EW posterior displays a clear peak at $\approx 400~\text{Å}$, with an extended tail to higher values. This peak occurs close to $100~\text{Å}$ for the \textsc{very early start} case. The former is still on the high end of the observed EW distribution, but is plausible (see for instance \cite{Ouchi2008}, which reports a few such LAEs), and the latter is typical of LAEs of similar $M_{\rm UV}$. Note, however, that the \textsc{very early start} model has a CMB optical depth of $\tau_\text{CMB}=0.085$, which is in significant tension with the Planck measurement of $\tau_\text{CMB}=0.058^{+0.006}_{-0.006}$ (Figure \ref{fig:compare_models})~\citep{Tristram2024}. Both of our models that fall within CMB optical depth constraints require JADES-GS-z13-1-LA to have fairly extreme Ly$\alpha$ emission properties. 

We further calculated the emission properties of JADES-GS-z13-1-LA using halos with $M_\text{UV}>-18$ to assess the effect of source clustering on our results. Specifically, we expect that the brightest galaxies in our volume will live close to the centres of the largest ionised regions, and thus have elevated IGM transmission relative to fainter galaxies. We find that using sightlines from fainter galaxies results in modestly reduced IGM transmission in all our models, which in turn requires the intrinsic Ly$\alpha$ emission properties of JADES-GS-z13-1-LA to be more extreme. As an example, using fainter halos with the \textsc{early start} model increases the inferred velocity offset by $300~\text{km/s}$ and the EW by $200~\text{Å}$. Because we know little about the environment of JADES-GS-z13-1-LA, there is no reason to prefer more or less isolated halos, so we assume that source clustering has a negligible effect on our analysis. As we will see in the next section, if the intrinsic properties of JADES-GS-z13-1-LA are more extreme than we infer, it would only strengthen our main conclusions, so this is a conservative assumption.

Our findings suggest that if JADES-GS-z13-1-LA is a star-forming galaxy with typical Ly$\alpha$ emission properties, it is likely telling us that reionisation was underway by $z = 13$. This motivates a more sophisticated analysis of whether or not this galaxy hosts an AGN, which we undertake next. 

\section{Estimating the probability of AGN activity}
\label{sec:likelihood}

Typically, distinguishing between an AGN and a purely star-forming galaxy (SFG) requires observations in the radio and X-ray bands~\citep{Padmanabhan2021}, SED fitting~\citep{DSilva2025}, or emission line diagnostics~(for instance, \cite{Maiolino2024}). Due to the limited spectroscopic data we have for JADES-GS-z13-1-LA, we cannot use these conventional methods to distinguish between the AGN and SFG scenarios. Therefore, in this section, we use a Bayesian approach to quantitatively assess whether the properties inferred for JADES-GS-z13-1-LA indicate that its Ly$\alpha$ emission is likely driven by AGN activity.

We denote the inferred properties of JADES-GS-z13-1-LA by $D$. Our probability space has three parameters: $\text{EW}$, $\Delta v$, and a category variable $c\in\{\text{SFG},\text{AGN}\}$ which indicates whether or not a galaxy hosts an $\text{AGN}$. For brevity, let $\theta$ denote the parameters $\text{EW}$ and $\Delta v$. Let $p(D|\theta,c)$ be the likelihood function, which is proportional to the posterior distribution shown in Figure \ref{fig:smallposteriors}. We suppose that this function is independent of $c$, so that $p(D|\theta,c)=p(D|\theta)$. To apply Bayes' rule, we first marginalise the likelihood over $\theta$ to compute \begin{equation} p(D|c)=\int p(D|\theta)p(\theta|c)~\text{d}\theta, \label{eq:marglike} \end{equation} where $p(\theta|c)$ is the prior PDF of $\theta$ given a specific value of $c$, which corresponds to the observed distributions of $\text{EW}$ and $\Delta v$ for SFGs and AGNs. We implement this integral as a weighted sum over the samples in the \textsc{dynesty} posterior.

We then apply Bayes' rule to compute the posterior probability \begin{equation} P(c|D)=\frac{p(D|c)}{p(D)}P(c),\end{equation} where $P(c)$ is the prior probability of $c$ and $p(D)=\sum_cp(D|c)P(c)$ is the marginalised likelihood or evidence. We compute this posterior for each of our three reionisation histories, obtaining a relationship between $P(\text{AGN}|D)$ and the ionised fraction at $z=13$.

\subsection{The observed distribution of LAEs}
\label{subsec:PDSFG}


We use the \textit{Fiducial} $\text{EW}$ and $\Delta v$ distributions given in Eqns. 3-4 of~\citet{Qin2024} to model the PDFs of EW and $\Delta v$ for the $z = 13$ galaxy population, which we denote as $p_\text{obs}(\text{EW})$ and $p_\text{obs}(\Delta v)$. These add redshift evolution to the $z\sim6$ distributions given in \citet{Mason2018a} upon which they are based. The original distribution is derived from the \citet{DeBarros2017} and \citet{Pentericci2018} samples, and the redshift-evolved distribution includes samples from \citet{Witstok2024a} and \citet{Tang2024b}.\footnote{This distribution is computed from an LAE sample that does not discriminate between SFGs and AGNs. However, this LAE sample is a sample of all LAEs, not just ultraluminous ones, so it should not have a significant AGN fraction~\citep{Calhau2020}. Any error from AGN contamination would move the distribution to higher $\text{EW}$, increasing the likelihood of the SFG scenario.} For the purpose of modelling probabilities, we use $M_\text{UV}=-18.7$~\citep{Witstok2024} and a dark matter halo mass of $1.06\cdot10^{10} M_{\odot}$, which we derive from our simulation's mass-luminosity relations, as derived via abundance matching (see \S\ref{subsec:FlexRT}). The former determines the $\text{EW}$ distribution and the latter determines the $\Delta v$ distribution~\citep{Qin2024}.

Setting $p(\text{EW},\Delta v|\text{SFG})=p_\text{obs}(\text{EW})p_\text{obs}(\Delta v)$ assumes that the $\text{EW}$ and $\Delta v$ are statistically independent. In reality, they are likely correlated~\citep{Erb2014}, albeit with significant intrinsic scatter. 
Part of this correlation is accounted for in the assumed $M_{\rm UV}$ dependence of the intrinsic PDF (see~\citet{Mason2018a}). Since our distributions are derived from galaxies in a small UV magnitude range, we do not expect $M_{\rm UV}$ dependence to introduce a significant correlation. However, some additional correlation may arise from Ly$\alpha$ RT effects within galaxies, which we are not able to account for here. As in~\citet{Qin2024}, we use our PDF of observed LAE properties to derive a PDF of intrinsic LAE properties by accounting for IGM attenuation of the $z\geq6$ LAE sample (from \cite{Tang2024b}, \cite{Witstok2024a}, and references therein), from which our observational $\text{EW}$ distribution is derived. To this end, we use the (properly normalised) distribution of \textit{intrinsic} $\text{EW}$ and $\Delta v$ for LAEs:
\begin{equation}
\label{eqn:pSFG}
    p(\text{EW},\Delta v|\text{SFG}) \propto p_\text{obs}(\text{EW}\cdot \mathcal{T}_6(\Delta v))p_\text{obs}(\Delta v).
\end{equation}
Here we introduce the attenuation factor $\mathcal{T}_6(\Delta v)$, which denotes the ratio between the observed EW and the intrinsic EW of a Ly$\alpha$ emission line with velocity offset $\Delta v$ emitted at $z\sim6$. Multiplying the intrinsic EW of an emission line by $\mathcal{T}_6(\Delta v)$ gives the corresponding observed EW. We compute $\mathcal{T}_6(\Delta v)$ from our simulations, so it additionally depends on the reionisation history. As a necessary simplification, we compute the function $\mathcal{T}_6$ using the average transmission curve for each reionisation history, and use the same function $\mathcal{T}_6$ for all sightlines in each reionisation history. Most sightlines at $z\sim 6$ have an attenuation factor of order unity, so this is not significantly different from computing attenuation factors for each individual sightline. Note that Equation \ref{eqn:pSFG} breaks the assumption of independence between the EW and $\Delta v$ PDFs.

\subsection{Modeling an AGN distribution}
\label{subsec:PDAGN}
While Ly$\alpha$ EW has been measured for statistical samples of quasars at $z \sim 5-6$~\citep{Banados2016,Gloudemans2022}, these objects are typically $5-10$ magnitudes brighter than JADES-GS-z13-1-LA. As such, we cannot directly estimate the intrinsic Ly$\alpha$ emission properties for AGN of comparable brightness, as we do for SFGs. 
Therefore, to estimate $p(\text{EW},\Delta v|\text{AGN})$, we assume that the shape of the EW distribution is the same for AGNs as for SFGs, but shifted higher by a factor proportional to the ratio of the mean ionising photon production rates of AGN and SFGs. This implicitly assumes that the gas in the ISM/CGM of galaxies being in photo-ionisation equilibrium, such that Ly$\alpha$ emission from recombinations is proportional to the local absorption of ionising photons. It also makes the assumption that the radiative transfer physics of an AGN host galaxy is identical to that of an SFG. A lack of data on AGN-powered LAEs forces us to use this simplified model, which is a simple way to implement the central assumption that AGNs produce more ionising photons than SFGs. 

Using this assumption, we estimate an $\text{EW}$ distribution for AGN-powered LAEs by horizontally rescaling the SFG $\text{EW}$ distribution by a factor of the ratio $$r_{\xi_\text{ion}}=\xi_\text{ion,AGN}/\xi_\text{ion,LAE}=2.88, $$ where $$\xi_\text{ion,AGN}\approx10^{25.9}~\text{Hz}~\text{erg}^{-1}$$ is the average ionising production efficiency of AGNs~\citep{Matsuoka2018} and $$\xi_\text{ion,LAE}\approx10^{25.4}~\text{Hz}~\text{erg}^{-1}$$ is the average ionising production efficiency of LAEs (\cite{Simmonds2023}, see also \cite{Saxena2024}). This amounts to setting 
\begin{equation}
    \label{pAGN}
    p(\text{EW},\Delta v|\text{AGN})\propto p(\text{EW}/r_{\xi_\text{ion}},\Delta v|\text{SFG}).
\end{equation}

JADES-GS-z13-1-LA is inferred to have $$\xi_\text{ion}\approx10^{26.5}~\text{Hz}~\text{erg}^{-1},$$ significantly greater than the average for either AGNs or LAEs~\citep{Witstok2024}. However, because AGNs generally have greater $\xi_\text{ion}$ than ordinary LAEs, this high $\xi_\text{ion}$ favours the AGN hypothesis. 
This approach quantifies our earlier suggestion that the extreme inferred properties of JADES-GS-z13-1-LA may indicate AGN activity.



\subsection{Results}


As our prior probability, $P(\text{AGN})$, we use the overall AGN fraction of galaxies in the JADES survey, $f_\text{AGN}=20 \pm 5\%$, as reported in~\citet{Scholtz2025}. This is the probability that a randomly chosen galaxy hosts an AGN, so it is the probability that JADES-GS-z13-1-LA hosts an AGN before accounting for its emission properties. $P(\text{SFG})$ is simply given by $1-P(\text{AGN})$. 


\begin{figure}[h!]
    \includegraphics[scale=.6]{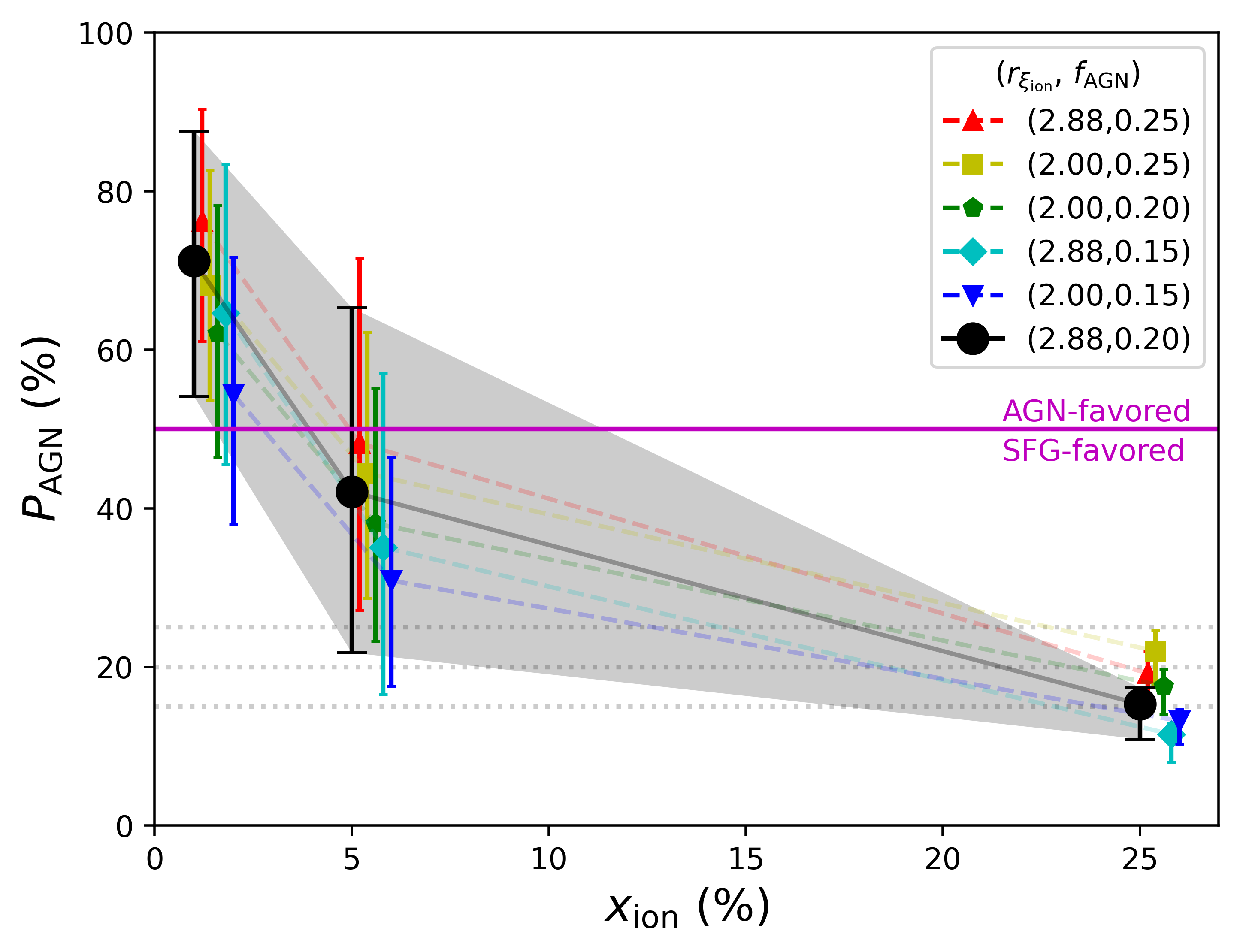}
    \caption{Values of $P(\text{AGN}|D)$, including uncertainties, for varying values of $r_{\xi_\text{ion}}$ and $f_\text{AGN}$. Our fiducial value, for $r_{\xi_\text{ion}}=2.88$ and $f_\text{AGN}=0.20$, is shown in black. We compare this to probabilities obtained with a lower $r_{\xi_\text{ion}}$ of $2.00$ to show how $r_{\xi_\text{ion}}$ affects the estimate. We also compare this to probabilities obtained with priors at the upper and lower 1$\sigma$ limits from~\citet{Scholtz2025}. These priors, along with our fiducial prior, are shown on the plot as dotted grey lines.}
    \label{fig:probplot}
\end{figure}

The dependence of the probability on IGM neutral fraction is shown in Figure \ref{fig:probplot}. The large black points show our fiducial values $P(\text{AGN}|D)$, and the magenta line denotes the boundary between a preference for the AGN and SFG scenarios. To show the importance of our assumptions in estimating these probabilities, we also plot results with varying values of $r_{\xi_\text{ion}}$ and $f_\text{AGN}$ in Figure \ref{fig:probplot}. A lower value of $r_{\xi_\text{ion}}=2.00$, which is suggested by some results which find a stronger upward redshift evolution of $\xi_\text{ion}$ for SFGs~\citep{Papovich2025}, decreases the estimated $P(\text{AGN}|D)$ by more than 10\%. However, even with lower values of $r_{\xi_\text{ion}}$ and $f_\text{AGN}$, our main conclusions hold. In particular, the \textsc{late start} model somewhat favours the AGN scenario even with the most conservative parameters (the blue downward triangles), and the \textsc{very early start} model strongly favours the SFG scenario even with the most optimistic parameters (the red upward triangles). The \textsc{early start} model has the most uncertainty, probably because (i) it is at a transitional point of reionisation where sightline to sightline (and halo to halo) variation is particularly high, and (ii) the probability is further away from the minimum (prior)\footnote{In this case, we generally expect $P(D|\text{}AGN)>P(D|\text{SFG})$, which implies $P(\text{AGN}|D)>P(\text{AGN})$.} and maximum (100\%) values that it can attain, so the distribution of probabilities is wider. 


Owing to the dearth of available data on high-redshift AGN-driven LAE populations and the general difficulty of detecting Type 2 AGNs, our result can only be considered a rough estimate. \citet{Calhau2020} notes that AGNs are systematically under-detected because obscured AGNs are often difficult to detect in the radio and X-ray bands. Therefore, since our calculation is based on empirically derived AGN fractions, it may be an underestimate.

If JADES-GS-z13-1-LA is not an AGN, our results (at face value) favour an ionised fraction in excess of a couple percent at $z = 13$. Recent observations tend to support a scenario in which the bulk of reionisation started relatively late at $z < 10$~\citep{Cain2024b}, disfavouring our \textsc{very early start} scenario. However, reionisation could still have a long, extended tail well above $z = 10$~\citep{Park2013}, or that the early stages of reionisation may have been non-monotonic, with a "bump" of ionisation of high redshifts~\citep{Cen2003,Ahn2021}. These types of scenarios often involve some early contribution to reionisation, at the $5-10\%$ level, from Pop. III stars~\citep{Tan2025}, and/or contributions to reionisation from mini-halos~\citep{Norman2018}. At face value, such early reionisation scenarios are generally disfavoured by constraints on the electron scattering optical depth $\tau_{\rm CMB}$ from the low-$\ell$ polarisation of the CMB measured by Planck~\citep{Wu2021c}. However, it has been recently noted that the latest DESI-DR2 data~\citep{DESIDR22025} may prefer a higher value of $\tau_{\rm CMB}$ in combination with the CMB\footnote{See also~\citet{Allali2025} for a similar type of result related to the Hubble tension. }~\citep{Sailer2025,Jhaveri2025} (although see~\citet{Cain2025} for difficulties with this scenario). Forthcoming CMB experiments such as LiteBIRD~\citep{Matsumara2014} will help resolve this question. 

\subsection{Corroborating the AGN scenario}

A few additional lines of evidence support the possibility that JADES-GS-z13-1-LA hosts an AGN. On their own, they are tentative evidence at best, but together with the relatively high estimated probability of the AGN scenario in both of our more plausible reionisation histories, they collectively paint a compelling picture of this possibility. 

As mentioned in \S\ref{subsec:spectrum} and discussed in~\citet{Witstok2024}, the UV turnover in the JADES-GS-z13-1-LA spectrum is best explained by a DLA with HI column density $\sim10^{23}~\text{cm}^{-2}$. This is consistent with the column density of the torus that surrounds an AGN~\citep{Peca2021}. This would account for both the obscuration of the AGN and the damping wing feature. AGN obscured by dense HI in their torus often display significant variability in HI column density on short timescales~\citep{Liu2017,Cox2025}. If follow-up observations of JADES-GS-z13-1-LA detected clear evidence of changes in $N_{\rm HI}$, this would be compelling evidence for the torus obscuration scenario. 

In \S\ref{subsec:spectrum}, we also report a tentative, potential detection of a CII* doublet emission line at $\lambda=1334,1335~\text{Å}$ (Figure \ref{fig:fit}). This potential line was not mentioned in~\citet{Witstok2024}, perhaps because it coincides with the peak of the DLA feature. This emission line is often detected in quasar spectra~\citep{VandenBerk2001,Maiolino2024}. On the other hand, this line is rarely detected in SFG spectra, and when it is detected in these spectra it is usually an absorption feature~\citep{Berg2022}. If higher-resolution followup spectroscopy confirms the presence of a strong CII* emission line, especially one without a resonant absorption feature, it would be compelling evidence that this galaxy hosts an AGN.

For JADES-GS-z13-1-LA, \citet{Witstok2024} reports a Ly$\alpha$ luminosity of $\sim10^{43.4}~\text{erg}~\text{s}^{-1}$, and we infer that the Ly$\alpha$ luminosity is $\gtrsim10^{43}~\text{erg}~\text{s}^{-1}$. At this luminosity, a relatively high fraction of LAEs host AGNs~\citep{Konno2016,Calhau2020}, although this fraction has not been determined beyond $z\sim6$. Additionally,~\citet{Baek2013} find that, in dark matter halos of masses near $\sim 10^{10}$ $M_{\odot}$, SFGs cannot produce the Ly$\alpha$ emission we observe for this galaxy, while Compton-thin AGNs (that is, AGNs with obscuring column densities of $\sim10^{22-24}~\text{cm}^{-2}$) can. This statement is related to our previous argument about emission features, but qualitatively corroborates it with independent results that link extreme Ly$\alpha$ emission and AGN activity.

\section{Conclusions}
\label{sec:conc}

We have performed detailed modelling of the spectrum of JADES-GS-z13-1-LA, focusing on the prominent Ly$\alpha$ emission feature and its potential implications for reionisation. In this context, we also investigated the possibility that the Ly$\alpha$ emission from JADES-GS-z13-1-LA is driven by an AGN. We find that the probability that JADES-GS-z13-1-LA hosts an AGN is $71\%$, $42\%$, and $15\%$ if the ionised fraction of the IGM is $< 1\%$, $\approx 5\%$, and $\approx 25\%$, respectively. These results reflect the fact that the inferred Ly$\alpha$ emission properties of JADES-GS-z13-1-LA are unlikely for star-forming galaxies unless reionisation is well-underway by $z = 13$. Because the the $25\%$ ionised fraction scenario is unlikely given current constraints from Planck, we conclude that JADES-GS-z13-1-LA is at least a good AGN candidate, and merits follow-up observations to ascertain whether or not it really hosts an AGN. Our stronger conclusions are conditioned on whether or not this galaxy hosts an AGN. 

If we confirm that JADES-GS-z13-1-LA hosts an AGN, then we can be reasonably confident that we are looking at an obscured AGN hiding behind a torus of mostly neutral hydrogen (as argued in~\cite{Witstok2024}). This AGN is probably a prodigious producer of LyC photons, perhaps carving out an ionised bubble around itself and causing extremely luminous Ly$\alpha$ emission in recombining gas. This hints at the existence of more such AGN-powered LAEs hiding in faint high-redshift galaxies~\citep{Fujimoto2024,Witstok2025b}; observing more of these may lend empirical credence to the possibility that AGNs contributed to the early stages of reionisation. If we find that JADES-GS-z13-1-LA does not host an AGN, then reionisation was probably underway by $z=13$, with $x_\text{ion}$ at least a couple percent. It is unlikely that a pure SFG could produce the intrinsic Ly$\alpha$ emission (with $\text{EW}\gtrsim1000~\text{Å}$) that we infer for this galaxy in the \textsc{late start} reionisation history. As pointed out by~\citet{Sailer2025,Jhaveri2025}, an early start to reionisation, and a high CMB optical depth, may help relieve recent tensions between DESI DR2 data and the CMB (although see~\cite{Cain2025}). 

Our findings suggest that follow-up observations of JADES-GS-z13-1-LA will be crucial to understand its nature. Therefore, follow-up observations focused on individual AGN-correlated emission lines, specifically NV$\lambda1240$ and the aforementioned CII* line, could help to constrain the AGN contribution to this galaxy's luminosity. Because it is already suggested by our data, the CII* line could be an especially cost-effective and promising AGN diagnostic for this galaxy.

As pointed out in~\citet{Witstok2024}, JADES-GS-z13-1-LA was detected in a relatively small survey area. Therefore, there are probably many more observable LAEs at $z\gtrsim10$ waiting to be discovered by JWST. Finding more of these LAEs and understanding their properties could be instrumental in understanding the evolution of high-redshift galaxies, constraining the history of reionisation, and determining what population of galaxies reionised the universe.

\paragraph{Acknowledgements}

The authors thank Hy Trac for providing access to his cosmological hydrodynamics code, acknowledge helpful conversations with Matthew McQuinn and Shabbir Shaikh, and thank Joris Witstok and the anonymous referee for helpful comments on the draft version of this manuscript. We also thank Kevin Croker for his help installing and running the \textsc{dynesty} package.  

\paragraph{Funding Statement}

JC, CC, and TC were supported by the Beus Center for Cosmic Foundations. RAW acknowledges support from NASA JWST Interdisciplinary Scientist grants
NAG5-12460, NNX14AN10G and 80NSSC18K0200 from GSFC. AD was supported by NSF AST-2045600 and JWSTAR02608.001-A. YZ acknowledges support from the NIRCam Science Team contract to the University of Arizona, NAS5-02015.

\paragraph{Competing Interests}

The authors are unaware of any competing interests. 

\paragraph{Data Availability Statement}

The data underlying this article will be shared upon reasonable request to the corresponding author. 

\paragraph{Ethical Standards}
The research meets all ethical guidelines, including adherence to the legal requirements of the study country.

\paragraph{Author Contributions}

JC did all major calculations and analysis and led the writing of the manuscript. CC provided project guidance, assistance with some analysis, and assistance in drafting and editing the manuscript. All other co-authors provided helpful comments and feedback that strengthened the analysis and presentation of results. 

\printendnotes

\defbibnote{preamble}{ }

\printbibliography[prenote={preamble}]

\appendix


\section{Posteriors for all parameters}
\label{app:posteriors}

In this appendix we provide a full corner plot for the \textsc{ early start} posterior described in \S\ref{sec:inference} (see Figure \ref{fig:smallposteriors}c).

\begin{figure}[b!]
    \includegraphics[scale=0.45]{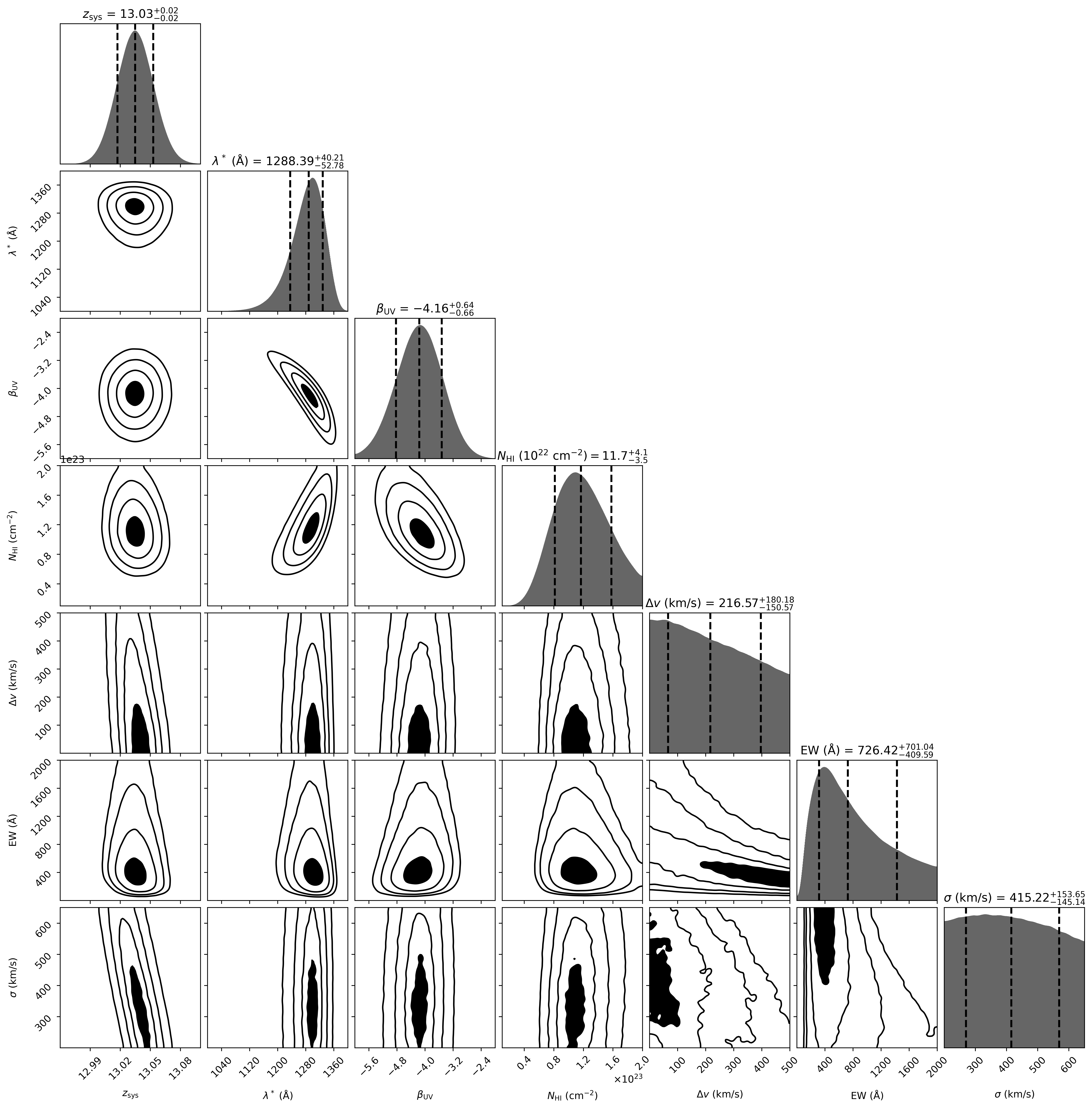}
    \caption{Full posteriors on all parameters listed in Table \ref{table:params} for the \textsc{early start} model.  }
    \label{fig:earlyfullcorner}
\end{figure}

\end{document}